\documentclass[12pt,preprint]{aastex}
\usepackage{graphicx}

\begin{document}

\title{The Molecular Disk in the Cloverleaf Quasar}

\author{S. Venturini}
\affil{Dept. of Physics \& Astronomy, State University of New York, Stony Brook, NY~11794-3800}
\email{Stefano.Venturini@sunysb.edu}
\author{P. M. Solomon}
\affil{Astronomy Program, State University of New York, Stony Brook, NY~11794}
\email{PSOLOMON@sbastk.ess.sunysb.edu}

\begin{abstract}
We propose a new interpretation for the CO emitting region of the Cloverleaf (H1413+1143), a gravitationally lensed QSO. We fit a two-galaxy lensing model directly to the IRAM CO(7-6) data rather than to the optical HST image and from the fit we infer that the CO(7-6) source is a disk-like structure with a characteristic radius of $785\;{\rm pc}$\footnote{In the currently widely accepted cosmology: cosmological constant $\Omega_{\Lambda}=0.7$, matter content $\Omega_{m}=0.3$, and Hubble constant $H_{0}=65\;{\rm km\;s^{-1}\;mpc}^{-1}$.}, a size similar to that of the CO emitting regions present in nearby starburst ultraluminous infrared galaxies. We therefore suggest that the Cloverleaf contains both an extended rotating molecular starburst disk and a central quasar.
\end{abstract}


\section{Introduction}
H1413+1143, also known as the Cloverleaf, is a broad absorption line QSO at a redshift of $z=2.55$ found by \citet{Duke}. It was subsequently identified as a lensed object with four bright image components (labeled from A to D as in Figure~\ref{fig1}a) by \citet{mag88}. 

The earliest indication that molecular emission lines could be present in the Cloverleaf spectrum came from the strong submillimeter dust continuum detected by \citet*{bar92} using the James Clerk Maxwell Telescope. Indeed, \citet{bar94} found CO(3-2) line emission using the Institut de Radioastronomie Millim\'etrique (IRAM) interferometer. This detection was soon followed by the detection of multiple CO transition lines by \citet{bar97} using the IRAM 30~m telescope and interferometer. Meanwhile, \citet*{wil95} obtained an interferometric map of the CO(3-2) emission using the Berkeley-Illinois-Maryland Array, but with too low spatial resolution to resolve the CO gas structure. Using the Owens Valley Millimeter Array (OVRO), \citet{yun97} obtained the first interferometric map of the Cloverleaf in which the CO(7-6) emission was partially resolved. \citet{all97} obtained a second map with sub-arcsecond resolution of the CO(7-6) line using the IRAM interferometer and \citet{kne98a} further enhanced it with additional data. The enhanced CO(7-6) map fully resolves the four different components of the Cloverleaf.

\citet{kay90} formulated the first lensing models. They described the lensing mass distribution of the two proposed models using respectively a singular isothermal elliptical galaxy and two singular isothermal spherical galaxies of equal masses. They constrained the models using an optical image and added radio features to the lensed source to model the Very Large Array images taken at the National Radio Astronomy Observatory. The resulting fit also suggested the hypothesis that microlensing effects might be present and important. Using these models, \citet{all97} estimated an upper bound of $1190 \; {\rm pc}$ for the characteristic radius of the CO(7-6) source.

\citet{yun97} modeled the lens using an elliptical potential with an external shear and used Hubble Space Telescope (HST) images to constrain the resulting lensing geometry. They constrained the size of the CO(7-6) source using the OVRO CO(7-6) data and estimated the corresponding characteristic radius to be $1420 \; {\rm pc}$. They also found that the magnification ratios of Keck K-band images reproduced the HST magnification ratios and interpreted this fact as an indication that microlensing was unimportant. However, \citet{ost97} found evidence that the D component of the lensed image is affected by microlensing. 

\citet{kne98a} also used the HST images to constrain the lensing geometry. They described the two proposed lensing mass distributions respectively by two truncated elliptical mass distributions (galaxy + dark halo) and a truncated elliptical mass distribution with an external shear (galaxy + cluster). They then used the enhanced IRAM interferometric CO(7-6) map to constrain the size of the CO(7-6) source and estimated the characteristic radii to be respectively $300 \; {\rm pc}$ and $100 \; {\rm pc}$.

In a subsequent observation, \citet*{kne98b} detected a possible candidate for the lensing galaxy on HST images. \citet{cha99} used the HST images and the position of the detected galaxy to constrain the lensing geometry. They modeled the lensing galaxy mass distribution using elliptical generalized power-law distributions. The proposed models used one single galaxy, one galaxy with external shear, and two galaxies as lenses. The two galaxies model gave the best fit and also the highest microlensing probability to the D component. \citet{cha01} observed polarization and intensity variations on HST optical images and suggested that microlensing effects are indeed present on the D component of the lensed image.

Unlike previous estimates of the size of the CO(7-6) source, we model both the source and the lens to obtain a best fit to the IRAM CO(7-6) image. We are thus able to derive an intrinsic size for the CO source using only CO data. Our best fit is able to reproduce the CO(7-6) map geometry as well as the brightness of the four image components. We find that the CO(7-6) source has a characteristic radius of $785 \; {\rm pc}$, similar to the size of nuclear disks present in local ultraluminous infrared galaxies (ULIRGs\footnote{ULIRGs have by definition high ${\rm 8 \! - \! 1000 \; \mu m}$ luminosities, i.e. ${\rm L_{IR}[8 \! - \! 1000 \; \mu m] \geq 10^{12}L_{\odot}}$.}). We therefore suggest that the Cloverleaf has a rotating molecular disk that harbors an extended starburst region, as is the case for nearby ULIRGs, and a central quasar.

This paper is organized in the following way: Section~2 describes the data we use, the model and the fitting procedures adopted, Section~3 presents our main results and comparisons with other models. Section~4 provides discussion, and in Section~5 we state our concluding remarks.

\section{The Model}
In our fitting procedure we use the velocity-integrated version of the IRAM CO(7-6) map that has been kindly provided to us by R.~Barvainis. We make two main assumptions in modeling the IRAM CO(7-6) data. First, we make the assumption that extinction along the line of sight is negligible. Second, we assume that microlensing is negligible in determining the overall geometry of the lensing system. These two assumptions together allow us to compare the map produced by our model to the IRAM CO(7-6) map.

\subsection{The Cosmology}
Angular-diameter distances, which relate angular separations to linear scales in the object frame, are one of the key quantities that enter the lens equation \citep*{sch92}:
\begin{equation}
\mbox{\boldmath$\eta$}=\frac{D_{s}}{D_{l}}\mbox{\boldmath$\zeta$}-D_{sl}\mbox{\boldmath$\alpha$}(\mbox{\boldmath$\zeta$}) \; ,
\end{equation}
where \mbox{\boldmath$\eta$} is the position vector of the QSO in the source plane and \mbox{\boldmath$\zeta$} the impact vector of the emitted light ray on the lens plane, \mbox{\boldmath$\alpha$} being the deflection angle. $D_{l}$ and $D_{s}$ are respectively the angular-diameter distances of the lens and the source from the observer, while $D_{sl}$ is the angular-diameter distance of the source from the lens (see Figure~\ref{fig2}). These distances are computed by integrating a Dyer-Roeder differential equation \citep{dye73} for the appropriate cosmology. For the cosmology we use ($\Omega_{\Lambda}=0.7$, $\Omega_{m}=0.3$), the equation is \citep{Kant}:
\begin{eqnarray}
& & \!\!\!\!\!\!\!\!\!\!\! (1+z)\left(\Omega_{m} (1+z)^{3} + \Omega_{\Lambda}\right) \frac{d^{2}D}{dz^{2}} + \left(\frac{7}{2}\Omega_{m}(1+z)^{3}+2\Omega_{\Lambda}\right)\frac{dD}{dz} +  \\ \nonumber
& & \mbox{$\;\;\;\;\;\;\;\;\;\;\;\;\;\;\;\;\;\;\;\;\;\;\;\;\;\;\;\;\;\;\;\;\;\;\;\;\;\;\;\;\;\;\;\;\;\;\;\;\;\;\;\;\;\;\;\;\;\;\;\;\;\;\;\;$} + \frac{3}{2} \: \alpha \Omega_{m} (1+z)^{2} D = 0 \; ,
\end{eqnarray}
where $D$ is the angular-diameter distance and $\alpha\in[0,1]$ is the so-called smoothness parameter. We do not estimate $\alpha$ but we analyze subsequently its impact on the source linear dimensions (see below). For an observer at $z_{1}$ and an object at $z_{2}$, the differential equation has the following boundary conditions:
\begin{eqnarray}
D(z_{1}) &=& 0   \\ \nonumber
\frac{dD}{dz}(z_{1})& =& \frac{c}{H_{0}} \, \frac{sign(z_{2}-z_{1})}{(1+z_{1}) \sqrt{\Omega_{m} (1+z_{1})^{3} + \Omega_{\Lambda} }} \; ,
\end{eqnarray}
where $D(z)$ is the angular-diameter distance of an object at redshift $z$ seen by the observer at redshift $z_{1}$.

\subsection{The Lens} 
The model assumes the presence of two lensing galaxies at a redshift of $z=1.55$. Each galaxy is modeled as an elliptical generalized power-law mass distribution. The resulting potential responsible of the lensing can be computed using the equations derived in \citet*{cha98}. The redshift of the lensing object candidate found by \citet{kne98b} is not known and the assumed value of $z=1.55$ is consistent with the redshift of observed absorbers in the line of sight \citep{mag88,tur88}. The angular-diameter distances to the lens and from the lens to the QSO depend on the redshift of the lensing object as seen previously but, since these distances affect only the linear size of the lensing galaxy and the critical density of the lens, the inferred properties of the CO source are not affected by the assumption made.

The elliptical generalized power-law mass distribution takes the following form \citep{cha98}:
\begin{equation}
k(\mbox{\boldmath$x$}) = k(r,\phi) = k_{0} \left \{ 1 + \left ( \frac{r}{r_{0}} \right )^{2} [ 1 + e \cos 2(\phi - \phi_{0})] \right  \}^{-\mu-1} \; ,
\end{equation}
where $k_{0}$ is the mass surface density at $r=0$ (i.e. the center of the distribution) in units of the critical mass density (defined as $c^{2}\,D_{s}/4\pi\,G\,D_{l}\,D_{sl}$), $e$ is the eccentricity, $\phi_{0}$ is the standard position angle, $r_{0}$ is the core radius and $\mu$ the radial index ($\mu=-1/2$ for an isothermal distribution). To these $5$ free parameters, we have to add $2$ more which describe the position of the center of the distribution on the lens plane. The total number of free parameters that describe the two galaxies distribution is therefore $14$. The only parameter that we do not minimize is the core radius. We arbitrarily keep it at a fixed value of $1.76\times 10^{-4}\;{\rm arcsec.}$ ($1.7\;{\rm pc}$ at the redshift of $z=1.55$) for both galaxies, a value similar to the one in \citet{cha99}. This is primarily due to the behavior of the mass density distribution for rays that are far from the core:
\begin{equation}
k(r,\phi) \approx k_{0} \, r_{0}^{2\mu+2} \left \{ r^{2} [ 1 + e \cos 2(\phi - \phi_{0}) ] \right \}^{-\mu-1} \; .
\end{equation}
Therefore, a change in the core radius $r_{0}$ can be very efficiently compensated by a change in the density $k_{0}$, leading to a flat (actually almost flat) direction in parameter space, an undesirable situation for a minimization routine. The smallness of the value of the core radius is dictated by the need of a rapid convergence in the numerical evaluation of the series expansion of the deflection angle given in \citet{cha98}.

The source is simply being modeled as a two dimensional Gaussian surface brightness distribution:
\begin{equation}
I(x,y) = I_{0} \, \exp \left \{ - \frac{x^{2}}{2\sigma_{x}^{2}} -  \frac{y^{2}}{2\sigma_{y}^{2}} \right \} \; ,
\end{equation} 
where $I_{0}$ is the central brightness, $\Delta x = \sqrt{2 \, ln{\, 2}} \; \sigma_{x}$ and $\Delta y = \sqrt{2\, ln{\, 2}} \; \sigma_{y}$ are the half width at half maximum (HWHM) of the Gaussian while $x$ and $y$ are the coordinates in a coordinate system rotated by a standard position angle $\psi_{0}$. Two more free parameters describe the position of the center of the brightness distribution on the source plane. The total number of free parameters for our model is then $17$, having kept the radial index of both galaxies equal.

\subsection{The Fitting Procedure}
To constrain the parameters of our model, we proceed by minimizing the following $\chi^{2}_{d.o.f.}$ defined as:
\begin{equation}
\chi^{2}_{d.o.f.} = \frac{\sum_{pix} [I_{data}(pix) - I_{model}(pix)]^{2}/\sigma^{2}}{N_{pix} - N_{par}} \; ,
\end{equation}
where the sum is carried over the pixels of the image. In our case the image is an array of $32\times32$ pixels, for a total of $N_{pix}=1024$. $N_{par}$ is an effective number of parameters that also takes into account the fact that image pixels within a beam-width are correlated. It is defined as the number of parameters of the model, i.e. $17$, multiplied by the number of pixels per beam-width, $17$ in the present case. Therefore we have $N_{par}=289$. $I_{data}(pix)$ and $I_{mod}(pix)$ are the intensity in the data and in the model maps of corresponding pixels. $\sigma$ is the noise level of the IRAM CO(7-6) map \citep{all97}. We determine the values of these $17$ parameters by minimizing such $\chi^{2}_{d.o.f}$. Due to the large amount of parameters, the choice of a minimization procedure fell on a simulated annealing algorithm such as the one described in \citet{nr}. As a remark, the number of data points included in the definition of the $\chi^{2}_{d.o.f.}$ is carefully chosen in order not to make our model over-parametrized but also to have a high signal to noise ratio on the overall image. In order to do so, the original map has been trimmed down to the smallest subset, shown in Figure~\ref{fig1}a, containing all the CO(7-6) emission from the Cloverleaf.

\section{Results}

\subsection{Main Results}
\begin{figure}
\begin{center}
\includegraphics[]{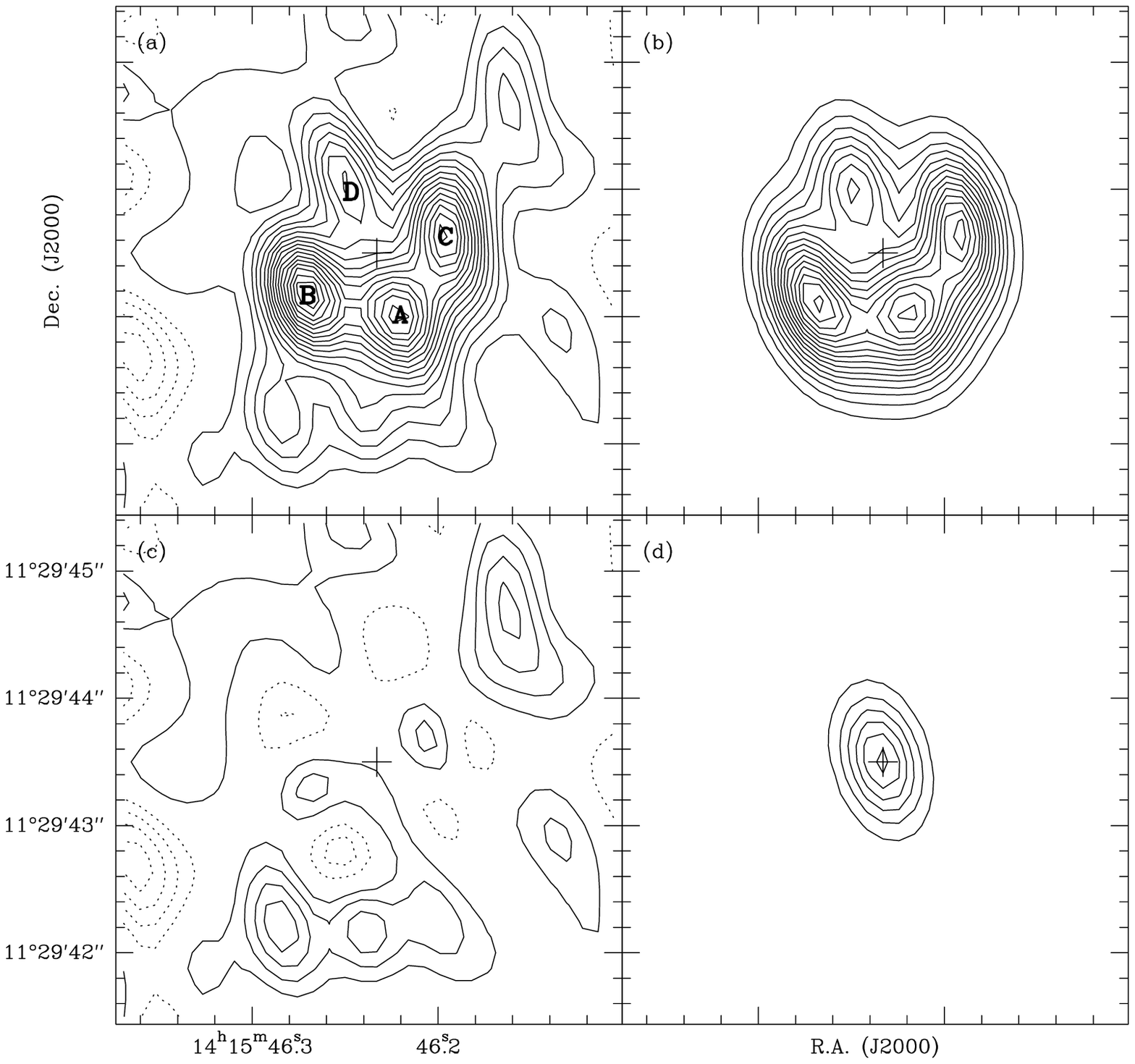}
\end{center}
\caption{(a) Integrated IRAM CO(7-6) map. (b) Model map. (c) Difference between data and model (a$-$b). (d) Unlensed CO source. In all the figures, the beam is $0.77''  \times 0.44''$ at a P.A. of $15^{\circ}$, with ${\rm T_{b}} / {\rm S} = 70 \; {\rm K \; Jy^{-1}}$, and the $1 \; \sigma$ contours are $0.55 \; {\rm Jy\; km \; s}^{-1} \; {\rm beam}^{-1}$.\label{fig1}}
\end{figure}

\begin{figure}
%\plotone{f2.eps}
%\includegraphics[scale=.815, trim= 0 140 0 90]{f2.eps}
\begin{center}
\includegraphics[width=8.2cm, trim= 0 140 0 90]{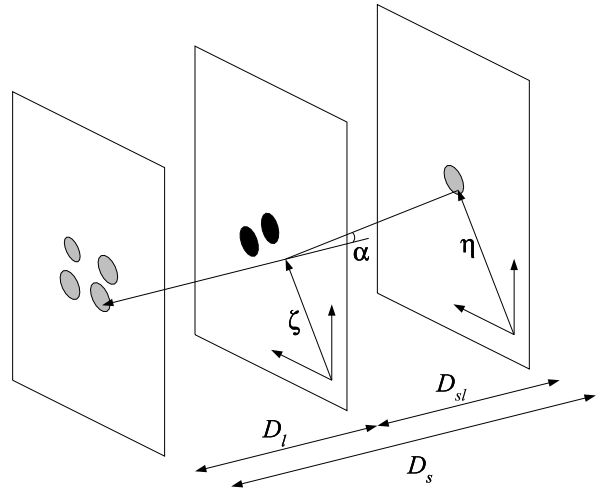}
\end{center}
\caption{A light ray from the source with position vector \mbox{\boldmath$\eta$} on the source plane  and impact parameter \mbox{\boldmath$\zeta$} on the lens plane is deflected by the deflection angle \mbox{\boldmath$\alpha$}. Also shown are the angular-diameter distances. \label{fig2}}
\end{figure}

By fitting our model directly the IRAM CO(7-6) data, we show that it is possible to constrain a lensing geometry even without the use of high resolution optical data such as HST images. Indeed, our model is able to reproduce the geometry as well as the brightness of the four images of the lensed QSO (see Figure~\ref{fig1}b). The good agreement with the data is confirmed by the $\chi^{2}_{d.o.f} = 2.8$. This good agreement is also clearly visible from Figure~\ref{fig1}c, the difference between data and model. From the same image we also notice that our simple model is not able to account for the weak extended emission. We list in Table~\ref{tab1} the parameters of the fit that describe the lensing galaxies as well as the parameters that describe the source. We notice that the derived position of one of the lensing galaxies is roughly consistent with the position of the candidate lensing galaxy found by \citet{kne98b}. We find that the CO(7-6) source has an effective size (HWHM) of $0.0794'' \times 0.0686''$. At the redshift of the Cloverleaf, the angular-diameter distance that relates angular to linear sizes does not vary more than $26 \%$ as $\alpha$, the smoothness parameter present in the Dyer-Roeder differential equation, ranges from $0$ to $1$. We fixed its value to $0.5$, making less than a $13 \%$ error in the linear dimensions of the CO source. With this value of $\alpha$, the angular size translates into a linear size of $785 \times 678 \;{\rm pc}$ in the adopted cosmology. The source surface brightness peak intensity of $11.8 \times 10^{3} \; {\rm K} \; {\rm km} \; {\rm s}^{-1}$ corresponds to a gas mass surface density of about $9.5 \times 10^{3} \; {\rm M_{\odot}/pc^{2}}$ if we assume that the conversion factor between CO luminosity \citep*{sol92} and gas mass has a value of $0.8$ \citep{dow98}. From our model we find that the CO(7-6) line emission from the QSO is magnified $11$ times.

\subsection{Comparison with Previous Models}
The first model we compare to is Model 1 from \citet{cha99}. The characteristics of the lenses are rather similar as far as the orientation of the mass isocontours are concerned ($33^{\circ}$ and $27^{\circ}$ compared to $24^{\circ}$ and $37^{\circ}$). The eccentricities are somewhat different ($0.852$ and $0.554$ compared to $0.39$ and $0.6$). The relative position of the two lensing galaxies are also similar, separated by a distance of $0.36$ arcseconds instead of $0.5$ and with a position angle of $146^{\circ}$ instead of $130^{\circ}$. However there is a big difference in the mass distribution since in our model the two galaxies have similar masses while the galaxies in Model 1 have a mass ratio of about $1:5$.

The subsequent models we consider are from \citet{kne98a}. These models differ substantially from our model since we use two galaxies as lens while \citet{kne98a} use a single galaxy, and a galaxy with an external shear. A marked difference occurs in the size of the CO(7-6) source. Our size of $785 \times 678\;{\rm pc}$ is much larger than the sizes of $300 \times 150 \;{\rm pc}$ and $100 \times 70\;{\rm pc}$ found by \citet{kne98a}. The magnification of $11$ that we find is also smaller than the magnification of $18$ and $30$ found by \citet{kne98a}.

The elliptical potential with external shear model by \citet{yun97} also differs substantially from our model. The size and magnification of the CO(7-6) source they derive ($1420 \; {\rm pc}$ and $10$ respectively) are consistent with our estimates. But it is difficult to assess how well their model maps the CO(7-6) images back to a single source on the lens plane since the OVRO CO(7-6) data (as well as its derived blueshifted and redshifted emission images) they base their analysis on is not able to resolve the four components of the Cloverleaf. Moreover, their model is clearly not able to map back all the redshifted CO(7-6) emission to a single source \citep[see Figure~2b in][]{yun97}.

Finally we consider the size estimate by \citet{all97}. They derive an upper limit of $1190 \;{\rm pc}$ for the CO(7-6) source radius, based on the size of the diamond caustic of the single elliptical isothermal galaxy model by \citet{kay90}. This estimate is consistent with ours, but it is based on a model that is not able to reproduce the brightness ratio of the different components of the optical image used to constrain it.

\section{Discussion: The Molecular Disk}
We list the intrinsic properties of the CO(7-6) source in Table~\ref{tab3} where we show the derived CO luminosity. From this table, it is clear that our model accurately accounts for the CO luminosity of the central part of the image containing the four bright image components. If we consider all the CO emission in the image at or above the $4 \; \sigma$ level, our model reproduces $\sim 95 \%$ of the luminosity. However, the model fails to reproduce the weak extended emission, surrounding the four image components, that contains $22\%$ of the total luminosity.

Given the presence of velocity structure in the IRAM CO(7-6) data and the size estimate from our model, we suggest that the CO(7-6) source might be a rotating disk-like structure with a characteristic radius of $785\;{\rm pc}$. Assuming a flat rotation curve, we isolated the part of the Gaussian source that contributes to the redshifted and blueshifted CO(7-6) emission images \citep[see Figure~1c and 1d in][]{all97} and produced the corresponding model images. Even with this rough approximation, the model images succeed in reproducing the geometry and the main features of the data images, further confirming our model.

An estimate of the dynamical mass can be made by assuming that the linewidth of the CO(7-6) emission line is due to rotation:
\begin{equation}
M_{dyn} \approx \frac{\Delta V^{2} \, R}{\sin^{2}(i) \; G} = \frac{2.6 \times 10^{10}}{\sin^{2}(i)} \; {\rm M}_{\odot} \; ,
\end{equation}
where $\Delta V$ is the half width at zero intensity of $375 \; {\rm km \; s}^{-1}$ \citep[see Figure~1 in][]{bar97}, $i$ is the inclination angle from face on and $R$ is the HWHM radius of $785 \; {\rm pc}$ found here. Recalling that for filled disks the CO luminosity traces the geometric mean of the dynamical mass and the gas mass \citep*{dow93}, we obtain:
\begin{equation}
M_{gas} = \alpha^{2} \, 3.0 \times 10^{10} \; \sin^{2}(i) \; {\rm M}_{\odot} \; ,
\end{equation}
where $\alpha$ is the conversion factor between $M_{gas}$ and $L'_{CO}$ for virialized clouds. Since the gas mass cannot exceed the dynamical mass, we have: $\alpha \leq 0.9 / \sin^{2}(i)$. For high inclination angles, this value of $\alpha$ becomes similar to the value of about $0.8$ found by \citet{dow98} for nearby ULIRGs. It becomes comparable to the value of $4.8$ found for our Galaxy \citep{sol91} only for unrealistically low inclination angles.

We can also have a rough estimate of the size of the Cloverleaf molecular disk by considering it a black body, made of optically thick dust. We can estimate the temperature of the dust using the $350 \; {\rm \mu m}$ (roughly $100 \; {\rm \mu m}$ in the rest frame) and $100 \; {\rm \mu m}$ (roughly $30 \; {\rm \mu m}$ in the rest frame) flux ratio:
\begin{equation}
\frac{F_{100} \, m_{350}}{F_{350} \, m_{100}} = \frac{\lambda_{350}^{3}}{\lambda_{100}^{3}} \; \frac{\exp \left( h \, c \, (z+1)/\lambda_{350} \, k \, T \right) - 1}{\exp \left( h \, c \, (z+1)/\lambda_{100} \, k \, T \right) - 1} \; ,
\end{equation}
where $F_{\lambda}$ (with $\lambda$ in $\mu m$) are the observed fluxes at the observed $\lambda$ wavelength \citep*[see][and references therein]{and99}, $m_{\lambda}$ the flux magnification, $z$ the redshift of the Cloverleaf and $T$ the black body temperature. We find that the black body temperature is roughly $135 \; {\rm K}$, having assumed equal magnification for both fluxes. We estimate a bolometric luminosity of roughly $1.1 \times 10^{14} \; {\rm L_{\odot}}$ from the spectral energy distribution \citep{gra96}, assuming an average magnification of $11$. This leads to an estimate for the black body size of the dust region:
\begin{equation}
R_{bb} = R_{\odot} \left( \frac{L}{L_{\odot}} \right)^{1/2} \left( \frac{T_{\odot}}{T} \right)^{2} \; ,
\end{equation}
where $L_{\odot}$, $R_{\odot}$, $T_{\odot}$ are the luminosity, radius and black body temperature of the sun. We find $R_{bb} \simeq 430 \; {\rm pc}$. Since the true size must be greater than the optically thick size of $430 \; {\rm pc}$, the value of $910 \; {\rm pc}$ found by \citet{gra96} and our value of $785 \; {\rm pc}$ for the CO(7-6) emitting region are acceptable while the estimates given by \citet{kne98a} of $300 \; {\rm pc}$ and $100 \; {\rm pc}$ for the CO(7-6) emitting region are much too small.

In Table~\ref{tab4} we compare the size and CO luminosity of the source for different cosmologies. The size of the source is similar to the size of nuclear molecular disks present in nearby ULIRGs \citep{dow98} as can be seen in Table~\ref{tab6}. We would like to stress that we are comparing different lines: the CO(7-6) one of the Cloverleaf and the CO(1-0) one of nearby ULIRGs. It is indeed true that the gas conditions and sizes of the emitting regions that these two lines probe are different. The size of the CO(7-6) emitting region gives a lower bound to the size of the lower excitation CO(1-0) emitting region. It is more difficult though to extend such simple considerations to the brightness temperature, given its complex dependence on the local gas conditions, and therefore to the CO luminosity. Unfortunately, with only the CO(7-6) high resolution interferometric data available, it is not possible to further investigate the observed similarities between these objects.

\section{Conclusions}
\begin{enumerate}
\item {\it Size of the CO(7-6) emitting region:} The characteristic size (HWHM) of the CO(7-6) line emitting region is $785 \; {\rm pc}$ in the adopted cosmology ($\Omega_{\Lambda}=0.7$, $\Omega_{m}=0.3$). The size of the lower excitation CO(1-0) line emitting region will be somewhat greater than this. Its size is therefore similar to the size of the nuclear molecular disks present in nearby ULIRGs.
\item {\it Molecular gas content:} Assuming that the source is indeed a disk-like structure with high inclination and assuming a conversion factor of $0.8$ between $M_{gas}$ and $L'_{CO}$, we find a molecular gas content of about $10^{10} \; {\rm M}_{\odot}$, consistent with the amount of molecular gas found in nearby ULIRGs.
\item {\it Starburst powered region:} The large size of the CO source (about $1 {\rm \; kpc}$) seems to rule out a scenario in which the molecular gas is concentrated in a very small region around the central AGN. This is consistent with the molecular gas being heated by a starburst, as in nearby ULIRGs. Therefore, two distinct sources coexist in the Cloverleaf: a central AGN and an extended starburst region.
\end{enumerate}

\acknowledgements
We would like to thank R.~Barvainis for having made the IRAM CO(7-6) integrated map available to us. We would also like to thank A.~Evans and J.~S.~Kim for helpful discussions.

\clearpage

\clearpage

\clearpage
\begin{deluxetable}{lc}
\tablecolumns{2}
\tablewidth{0pt}
\tablecaption{Model Parameters \tablenotemark{a} \label{tab1}}
\startdata
\cutinhead{Lens Parameters}
\sidehead{Galaxy 1}
$k_{0}\;\;(10^{6}\;{\rm M}_{\odot}\;{\rm pc}^{-2})$ & $18.3 \pm 0.4$ \\
$e$ & \phn $0.554 \pm 0.005$ \\
P.A.\tablenotemark{b}$\;\;\;({\rm deg.})$ & $27.0 \pm 0.4$ \\
$x_{G},y_{G}$\tablenotemark{c}$\;\;\;({\rm 10^{-2}\; arcsec.}$) & $20.8 \pm 0.3,-14.7 \pm 0.6$ \\
$\mu$\tablenotemark{d} & $-0.381 \pm 0.002$ \\
\sidehead{Galaxy 2}
$k_{0}\;\;(10^{6}\; {\rm M}_{\odot}\;{\rm pc}^{-2})$ & $12.7 \pm 0.1$ \\
$e$ & \phn $0.852 \pm 0.005$ \\
P.A.\tablenotemark{b}$\;\;\;({\rm deg.})$ & $33.0 \pm 0.5$ \\
$x_{G},y_{G}$\tablenotemark{c}$\;\;\;({\rm 10^{-2}\; arcsec.})$ & $0.69 \pm 0.07,15.3 \pm 0.2$ \\
$\mu$\tablenotemark{d} & $-0.381 \pm 0.002$ \\
\cutinhead{Source Parameters}
$I_{0}\;\;(10^{3} \; {\rm K} \; {\rm km} \; {\rm s}^{-1})$ & $11.8 \pm 0.2$\\
$\Delta x,\Delta y$ (${\rm 10^{-2}\; arcsec.}$) & $ 7.94 \pm 0.04, 6.86 \pm 0.04$ \tablenotemark{e} \\
P.A.\tablenotemark{b}$\;\;\;({\rm deg.})$ & $55 \pm 1$ \\
$x_{S},y_{S}$\tablenotemark{c}$\;\;\;({\rm 10^{-2}\; arcsec.})$ & $7.56 \pm 0.05, -5.07 \pm 0.05$ 
\enddata
\tablenotetext{a}{Error is computed as $\chi^{2}_{min} + 1.0$.}
\tablenotetext{b}{Position angle is East of North.}
\tablenotetext{c}{Offsets are with respect to the center of the IRAM CO(7-6) map at $14^{{\rm h}}15^{{\rm m}}46^{{\rm s}}.233$ RA and $11^{\circ}29'43''.50$ DEC, 2000 epoch.}
\tablenotetext{d}{The radial index is kept equal for both galaxies.}
\tablenotetext{e}{Half width at half power. In the adopted cosmology the corresponding linear sizes are $(785\pm5$, $678\pm5) \; {\rm pc}$.}
\end{deluxetable}

\clearpage
\begin{deluxetable}{lcc}
\tablecolumns{3}
\tablewidth{0pt}
\tablecaption{Source Properties \label{tab3}}
\tablehead{\colhead{} & \colhead{$L'_{CO}$} & \colhead{$L_{CO}$}\\
\colhead{} & \colhead{$10^{10} \; {\rm K \; km \; s^{-1} \; pc}^{2}$} & \colhead{$10^{8} \, {\rm L}_{\odot}$}}
\startdata
\cutinhead{Total Luminosity}
Data & $40 \pm  1$ & $68 \pm 3$\\
Model Lensed & $31 \pm 1$ & $53 \pm 2$\\
Model Unlensed & \phn $2.8 \pm 0.1$ & \phn $4.8 \pm 0.2$ \\
\cutinhead{Central Luminosity \tablenotemark{a}}
Data & $29 \pm  1$ & $49 \pm 2$\\
Model Lensed & $28 \pm 1$ & $47 \pm 1$
\enddata
\tablenotetext{a}{Within the $4 \; \sigma$ contour in the IRAM CO(7-6) map, i.e. including the four image components and excluding the weak extended emission.}
\end{deluxetable}

\clearpage
\begin{deluxetable}{cccc}
\tablecolumns{4} 
\tablewidth{0pt}
\tablecaption{Source Properties in Different Cosmologies \label{tab4}}
\tablehead{\colhead{$\alpha$} & \colhead{$L'_{CO}$ Data} & \colhead{$L'_{CO}$ Model Unlensed} & \colhead{$(\Delta x,\Delta y)\tablenotemark{a}$} \\
\colhead{} & \colhead{$10^{10} \; {\rm K \; km \; s^{-1} \; pc}^{2}$} & \colhead{$10^{10} \; {\rm K \; km \;s^{-1} \; pc}^{2}$} & \colhead{pc}}
\startdata
\cutinhead{$\Omega_{m}=0.3, \Omega_{\Lambda}=0.7, H_{0}=65\;{\rm km\;s^{-1}\;mpc}^{-1}$}
$0.0$ & $51$ & $3.6$ & $(891,770)$ \\
$0.5$ & $40$ & $2.8$ & $(785,678)$ \\
$1.0$ & $30$ & $2.1$ & $(687,594)$ \\
\cutinhead{$\Omega_{m}=1.0, \Omega_{\Lambda}=0.0, H_{0}=75\;{\rm km\;s^{-1}\;mpc}^{-1}$}
$0.0$ & $22$ & $1.6$ & $(590,509)$ \\
$0.5$ & $16$ & $1.1$ & $(493,426)$ \\
$1.0$ & $11$ & $0.7$ & $(407,351)$ \\
\enddata
\tablenotetext{a}{Half width at half power (HWHM).}
\end{deluxetable}

\clearpage
\begin{deluxetable}{lcccccc}
\tablewidth{0pt}
\tablecaption{Sizes of CO Emitting Regions in ULIRGs and High $z$ CO Sources\label{tab6}}
\tablehead{\colhead{} & \colhead{z} & \colhead{CO Line} & \colhead{Radius\tablenotemark{a}} & \colhead{True $T_{b}$} & \colhead{$L'_{CO}$\tablenotemark{b}} & \colhead{Reference}\\
\colhead{} & \colhead{} & \colhead{} & \colhead{pc} & K  & \colhead{$10^{9} \;{\rm K\;km\;s^{-1}\;pc}^{2}$}&}
\startdata
Arp 220 & $0.02$ & $1-0$ & \phn $560$ & $29$  & \phn $7.9$ & 1\\ 
Mrk 231 & $0.04$ & $1-0$ & \phn $540$ & $56$  & \phn $7.0$ & 1\\ 
Mrk 273 & $0.04$ & $1-0$ & \phn $460$ & $36$  & \phn $7.4$ & 1\\
VII Zw 31 & $0.05$ & $1-0$ & $1300$ & $18$  & $12.0$ & 1\\
IRAS 10214+4724\tablenotemark{c} & $2.29$ & $3-2$ & \phn $540$ & $40$  & $11.0$ & 2\\
 & & $6-5$ & \phn $540$ & $23$ & \phn $6.4$ & 2\\
Cloverleaf & $2.55$ & $7-6$ & \phn $785$ & $32$ & $28.0$ & 3\\
SMM J14011+0252 & $2.56$ & $3-2$ & \phn $410$ & $35$  & \phn $5.7$ & 4\\
 & & $7-6$ & \phn $480$ & \phn $7$  & \phn $1.3$ & 4\\
APM08279+5255\tablenotemark{d} & $3.91$ & $1-0$ & 1160 & 16 & 35.0 & 5 
\enddata
\tablenotetext{a}{Half power radius (HWHM).}
\tablenotetext{b}{Intrinsic luminosity.}
\tablenotetext{c}{Assuming a magnification of 15.}
\tablenotetext{d}{Assuming a magnification of 7.}
\tablerefs{(1)\citet{dow98}; (2)\citet{sol92}; (3)This article; (4)\citet{dow02}; (5)\citet{lew02}.}
\end{deluxetable}

\end{document}